\documentclass[epj]{svjour}
\usepackage{graphics}
\begin{document}
\title{A Short Review on Jet Identification}

\author{Sevil Salur\inst{1}\fnmsep\thanks{\email{ssalur@lbl.gov}} for the STAR Collaboration}
\institute{Lawrence Berkeley National Laboratory, 1 Cyclotron Road MS-70R0319, Berkeley, CA 94720}

%}                     % Do not remove
%
\offprints{}          % Insert a name or remove this line
\date{Received: date / Revised version: date}
\abstract{ Jets can be used to probe the physical properties of the high energy density matter  created in collisions at the Relativistic Heavy Ion Collider (RHIC). Measurements of  strong suppression of inclusive hadron distributions and di-hadron correlations at high $p_{T}$ have already provided evidence for partonic energy loss. However, these measurements suffer from well-known geometric biases due to the competition of energy loss and fragmentation.  These biases can be avoided if the jets are reconstructed independently of their fragmentation details - quenched or unquenched.  In this paper, we discuss modern jet reconstruction algorithms (cone and sequential recombination) and their corresponding background subtraction techniques  required by the high multiplicities of  heavy ion collisions. We review recent results from the STAR experiment at RHIC on direct jet reconstruction in central Au+Au collisions at $\sqrt {s_{NN} }= 200$ GeV. 
 % end of PACS codes
} %end of abstract
\maketitle
\section{Introduction}
\label{intro}

A highly collimated ``spray'' of particles also known as ``jets'' is produced in high energy collisions.   Hadronic jets are the experimental signatures of quarks and gluons (partons) and they reflect the underlying parton kinematics \cite{drell,cabibbo,bjorken,weinberg}. Cross section measurements of jets are performed at many hadronic and leptonic colliders to check in detail perturbative QCD (pQCD) calculations, to help determine parton distribution functions and to look for new physics. The inclusive jet cross section at Tevatron is measured very precisely over 20 orders of magnitude and is found to be in a very good agreement with the NLO pQCD calculations using CTEQ 6.1 parton distribution functions \cite{cdfall,cdf}. The robustness of the theoretical calculations on jet cross sections in $p+\bar{p}$ collisions motivates the use of jets as direct probes of partonic energy loss in dense matter generated in ultra-relativistic heavy ion 
collisions at  RHIC and in near future at the LHC \cite{starpt,phenixpt}.

Due to the limited pseudo-rapidity ($\eta$) and azimuthal $\phi$ coverage of the electromagnetic calorimeters of RHIC experiments during
the first 6 years of RHIC operation, the background determination needed
for direct jet reconstruction was not possible.
%During the first 6 years of RHIC operations, due to the limited $\eta$ and $\phi$ coverage of the electromagnetic calorimeters of RHIC experiments, the background determinations needed for direct jet reconstruction were not possible.  
Instead other observables such as the strong suppression of inclusive hadron distributions and di-hadron correlations at high $p_{T}$ were measured.   However, such measurements suffer from geometric biases due to the competition of energy loss and fragmentation - the leading particle spectrum is dominated by relatively low energy jets that happen to lose little energy in the medium and fragment into higher $p_{T}$ particles \cite{baier}. These biases can be removed if the reconstructed partonic kinematics is independent of whether the fragmentation is modified by the medium or not. An unbiased jet reconstruction measurement in heavy ion collisions would give access to the full spectrum of fragmentation topologies without geometric biases, enabling full exploration of quenching dynamics. In addition, fully reconstructed jets allow the measurement of qualitatively new observables such as jet shapes, fragmentation functions, and energy flow.   
Since 2006, the STAR barrel electromagnetic calorimeter (BEMC) with full azimuthal coverage and unit  pseudorapidity acceptance is operational. 
%In 2007 full azimuthal coverage at unit mid rapidity of the electromagnetic calorimeter is available for STAR experiment.  
This enables the study of the underlying event background required for full jet reconstruction in heavy ion collisions for the first time at RHIC. The experimental results discussed in this article were presented for the first time during the Hard Probes 2008 meeting \cite{me} for the first direct measurement of jets and \cite{jor} for the accompanying measurement of  jet fragmentation studies in heavy ion collisions.

\section{Jet Reconstruction Algorithms}\label{techno}  

During the last 20 years, various jet reconstruction algorithms were developed to combine measured particles into jets.    For a detailed overview of jet algorithms in high energy collisions, see \cite{blazey,salamtalk,catchment,jets} and references therein.  The primary requirement for jet algorithms is low sensitivity to hadronization, radiation and splitting.  They should also be defined equally at hadron and parton level. This was suggested by Weinberg 30 years ago \cite{weinberg}. According to him  ``Quark and gluon jets can be compared to detector jets, if jet algorithms respect collinear and infrared safety.''  

Full jet reconstruction in heavy ion collisions is a new frontier.  As in leptonic and hadronic collisions, in heavy ion collisions the chosen jet reconstruction method should also be theoretically and experimentally consistent.  As the experimental energy determination degrades the resolution somewhat with the detectors, the algorithm should also aim to minimize resolution effects from unrelated sources such as the underlying event. The expected increase in Large Hadron Collider  (LHC) luminosities (20 to 200 collisions in a detector) requires that the traditional jet algorithms have to be improved with underlying event subtraction techniques for $p+p$ collisions  to resolve events for pile up.  These improved techniques can also be used in heavy ion environments (Au+Au or Pb+Pb)  where the background subtraction is required due to large multiplicities of produced particles \cite{catchment,pileupcacciari}.

\begin{figure}[h!]
\begin{center}
\resizebox{0.45\textwidth}{!}{%
	\includegraphics{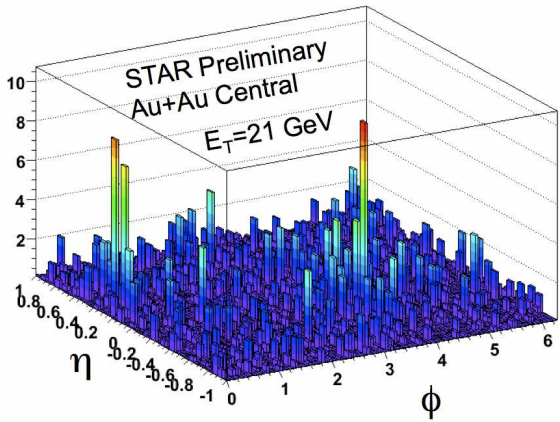} 
}	
\end{center}
\caption{21 GeV di-jet reconstructed from a single event with a combined transverse momentum and energy for charged and neutral particles per grid cell in the $\eta$ and $\phi$ plane from 0-20\% most central Au+Au collisions \cite{jor}.}
\label{fig:dijets}       % Give a unique label
\end{figure}

The algorithm should also be detector independent i.e., allow the combination of particles detected in various detectors.   The results discussed in this article are from jets that  are reconstructed by combining the neutral energy from the BEMC and charged particles from the Time Projection Chamber (TPC) of the STAR experiment.  STAR's TPC and BEMC detectors cover full azimuth ($0 < \phi <2 \pi$)  and mid rapidity ($ -1< \eta <1 $) of the events.   The Figure~\ref{fig:dijets} shows an example of a reconstructed  di-jet event in the STAR experiment.

Corrections for double-counting of energy due to hadronic energy deposition in the BEMC and to electrons are applied. Two kinds of jet reconstruction algorithms are utillized; seeded cone (leading order high seed cone (LOHSC)) and sequential recombination ($\rm k_{T}$ and Cambridge/Aachen). In the following, we briefly discuss these two algorithms and the corresponding underlying event subtractions.

\subsection{Cone Algorithms}\label{cone}

\begin{figure}[h!]
\begin{center}
\resizebox{0.35\textwidth}{!}{%
	\includegraphics{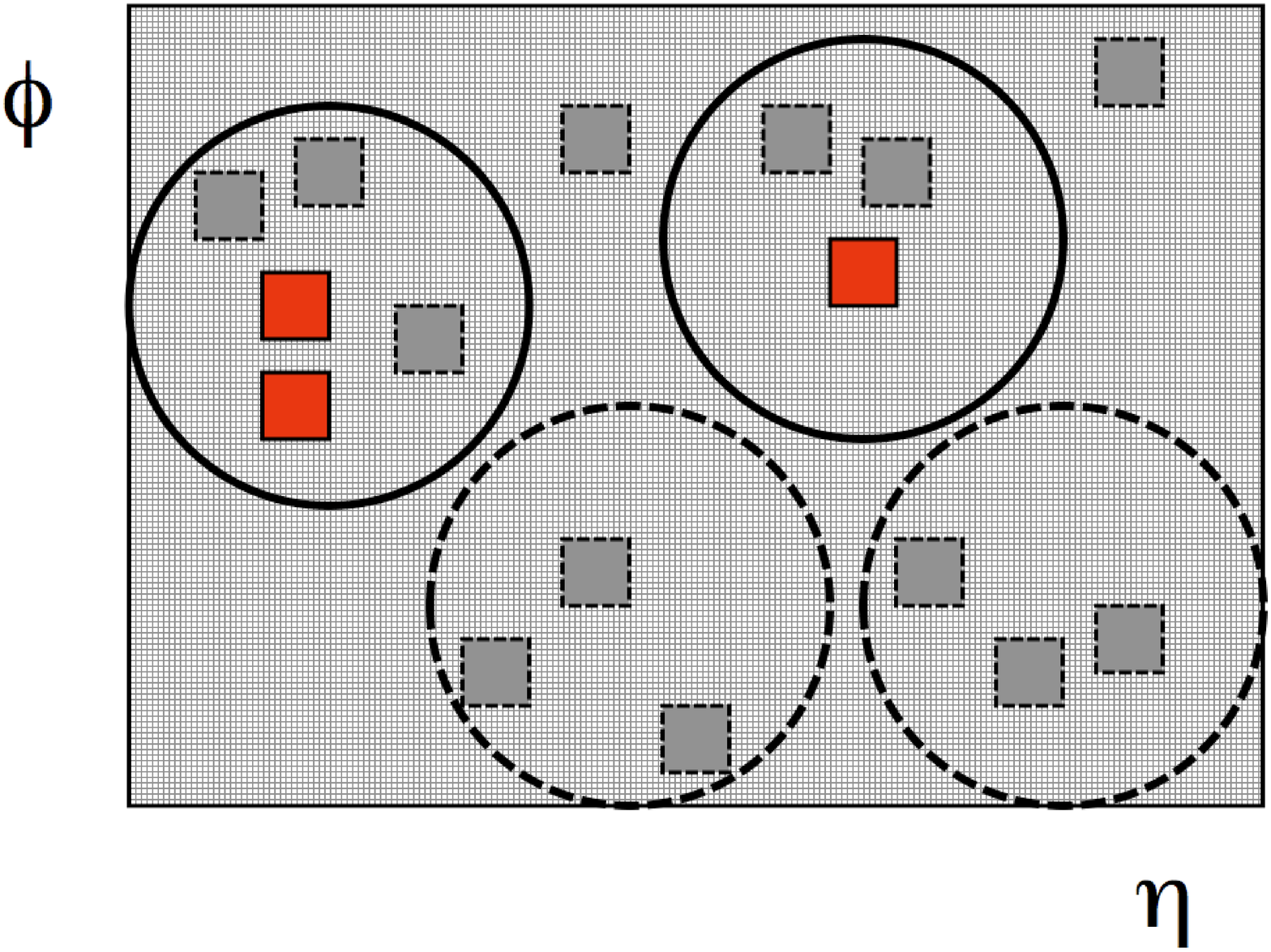} 
}	
\end{center}
\caption{The transverse momentum of tracks 
and/or deposition of energy in calorimeter towers are represented as the solid squares in $\eta$ and $\phi$ space. Red squares defined as seeds are the ones above a given threshold. Jet cones are solid circles around the seeds. All the energy of the particles is added for the given cone around the seed particles to estimate the jet energy.  The background in the jet energy is estimated with the average of the total energy in cones without seeds (shown in dashed circles) and is subtracted on an event-by-event basis. }
\label{fig:grid}       % Give a unique label
\end{figure}

The cone algorithms have been used as a primary tool to identify jets at hadron colliders since the early 1980s. This algorithm is based on the picture that a jet consists of a large amount of hadronic energy in a small angular region. Therefore, the main method is to combine particles in $\eta - \phi $ space with their neighbors within a cone of radius R ($R=\sqrt{ \Delta \phi ^{2}+ \Delta \eta^{2} }$).   This is illustrated schematically in Figure~\ref{fig:grid}. 

To optimize the search and effectiveness of jet finding,  these algorithms use splitting, merging, and iteration steps in the events of leptonic and hadronic collisions. However, to avoid instabilities in cone-finding due to large heavy ion background, we use a simple seeded cone without iteration or split-merging steps, with cone radius $R=0.4$ and minimum seed of 4.6 GeV.  The choice of the relatively small cone size is to suppress the underlying heavy ion background \cite{sarah,joern}.  In $p+p$ collisions $\sim 80$\% of the jet energy is observed to be within R$\sim0.3$ for 50 GeV jets in the Tevatron data \cite{cdf}.  However, broadening of the jet fragmentation due to quenching in the medium formed in heavy ion collisions may reduce the fraction of the measured energy in a given cone size and needs further exploration. To reduce the heavy ion background, the minimum accepted transverse momentum of charged particles, and the transverse energy of the calorimeter cells ($p_{T}^{cut}$) is  varied between 0.1 to 2 GeV.  This threshold cut does not remove all the background contamination on the jet energy and additional subtraction is needed. As presented in Figure~\ref{fig:grid} schematically,  the residual background is corrected based on the out-of-cone energy for the same $p_{T}^{cut}$, averaged over the STAR acceptance but measured on an event-by-event basis, and scaled to the cone area. 

A recently developed seedless infrared-safe cone algorithm (SISCone) \cite{sis} resolves most of the ambiguities associated with the various cone algorithms. The SISCone algorithm is already used in $p+p$ collisions at $\sqrt{s}=200$ GeV and the first results can be found in \cite{elena}.

%\begin{figure}[ht!]
%\begin{center}
%$\begin{array}{cc}
%\includegraphics[height=6.0cm,clip]{rcone07} &
%\includegraphics[height=6.0cm,clip]{rcone0704compare} \\
%\mbox{\bf (a) }&\mbox{\bf (b)}
%\end{array}$

%\caption[]{Comparison of the total energy {\bf(a) }for the two $p_{T}$ cuts (0.1 and 1 GeV) for a fixed $R=0.7$ and {\bf(b)} with two cone sizes ($R=0.4$ and $R=0.7$) for a fixed $p_{T}$ cut.  For these plots Hijing events are generated and separated into most central and peripheral Au+Au collisions at $\sqrt{s_{NN}} =200$ GeV. 
%}
%\label{fig1}
%\end{center}
%\end{figure} 

\subsection{Sequential Recombination Algorithms}\label{kt}

\begin{figure}[h!]
\begin{center}
\resizebox{0.20\textwidth}{!}{%
	\includegraphics{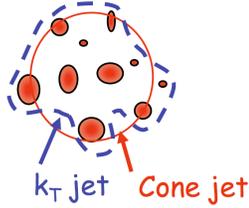} 
}	
\end{center}
\caption{A schematic comparison of the jet area for cone and $\rm k_{T}$ type algorithms.   }
\label{fig:ktconearea}       % Give a unique label
\end{figure}

The sequential recombination algorithms  have been used extensively in the Tevatron as they are collinear and infrared safe \cite{cdf,kt,ktref}.  In these types of algorithms, arbitrarily shaped jets are allowed to follow the energy flow resulting in less bias on the reconstructed jet shape than with cone algorithms \cite{catchment}.   Figure~\ref{fig:ktconearea} represents a schematic comparison of the jet areas for cone and $\rm k_{T}$ type algorithms.  

The sequential recombination algorithms  combine objects in relative to the closeness of their $p_{T}$. Particles are merged into a new cluster via successive pair-wise recombination.  Algorithmic details can be found in \cite{cdf,kt} and references therein.   The $FastJet$ code package for sequential recombination algorithms was used for the STAR analyses in $p+p$ and $Au+Au$ collisions \cite{me,jor,elena,fastjet,antikt}. This package includes $\rm k_{T}$, Cambridge/Aachen (CAMB), anti-$\rm k_{T}$, and an interface to external jet finders such as Seedless Infrared Safe Cone (SisCone)  via a plugin mechanism \cite{sis}.  For infrared and collinear safe algorithms  an active area ($A_{j}$) of each jet is estimated by filling an event with many very soft particles and then counting how many are clustered into a given jet.  If the underlying event is distributed uniformly in $\eta$ and $\phi$ then this noise density can be subtracted from the measured jet energy on an event-by-event basis to correct for the background energy underlying the jet. In simulations, this correction is observed to recover most of the jet energy when they are reconstructed in pile up and heavy ion backgrounds \cite{catchment}.  The $\rm k_{T}$, Cambridge/Aachen and anti-$\rm k_{T}$ algorithms are all based on the same sequential recombination algorithm, but they differ in the distance measure that is used to group particles in to jets. 

\begin{figure}[h!]
\begin{center}
\resizebox{0.40\textwidth}{!}{%
	\includegraphics{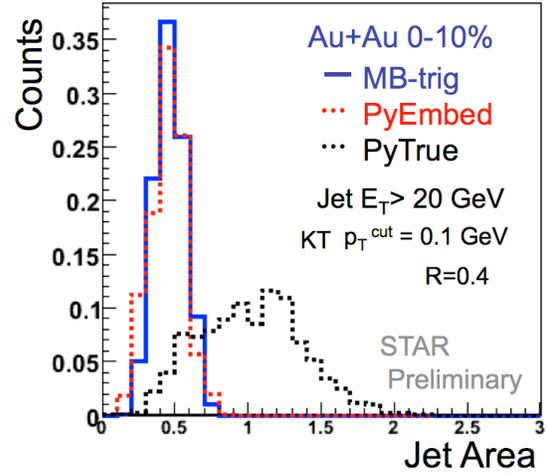} 
}	
\end{center}
\caption{Jet area from $\rm k_{T}$ algorithm is reconstructed utilizing the $FastJet$ code package \cite{catchment} for real jets in minimum bias triggered 0-10\% central $Au+Au$ collision (MB-trig),  in PYTHIA isolated jet events embedded in real central $Au+Au$ events (PyEmbed) and in PYTHIA isolated jet events (PyTrue). }
\label{fig:area}       % Give a unique label
\end{figure}

Figure \ref{fig:area} shows the jet-area from  $\rm k_{T}$  algorithm from $FastJet$ code for 0-10\% central Au+Au events (MB-Trig), compared to generated PYTHIA jets
(PyTrue) and PYTHIA jets embedded in heavy ion background (PyEmbed, see next section for details) \cite{me}.  PYTHIA jets embedded in real Au+Au background events are observed to have the same area as jets from real Au+Au events.  See the solid blue and dashed red histograms in Figure~\ref{fig:area}.  The reduction in the MB-Trig jet area relative  to PyTrue is well understood for sequential recombination algorithms on theoretical grounds \cite{catchment}.

\section{Jet Reconstruction Analysis}

\subsection{Event Selection and Terminology}

This analysis utilizes events of Au+Au collisions at $\sqrt{s_{NN}} = 200$ GeV recorded by STAR. Using the multiplicity measurement performed with the STAR TPC, only the most central (0-10\%) Au+Au collisions are selected. Two event sets were analyzed, based on fast on-line trigger configurations: \\
(i) MB-Trig, (minimum-bias trigger) utilizing the coincidence between the two calorimeters at beam rapidity (Zero Degree Calorimeters) with signals in each greater than $\sim40\%$ of the most probable amplitude for a single neutron, and \\
(ii) HT-Trig (high-tower trigger) that satisfies the  MB-Trig conditions and the additional requirement of 2-tower EMC clusters having at least a 7.5 GeV energy deposition. 

 %To be able to use the HT-Trig data for the inclusive jet spectrum, the bias due to the increased high $p_{T}$ data-taking is required to be corrected.
 
Three million MB-Trig events, corresponding to 300 thousand 0-10\% most central events are used for this study. A total of 80 million MB-Trig events were recorded by STAR during the 2007 $Au+Au$ run, but only the event set used have been fully reconstructed off-line. The HT-Trig is designed to enhance the recorded rate of high $p_{T}$ photons and electrons. It may also serve to enhance the recorded rate of jets. The HT-Trig data-set has been reconstructed in its entirety, corresponding to 500 $\rm \mu b^{-1}$.

In order to assess jet reconstruction energy resolution, background subtraction, efficiency and acceptance,  Monte-Carlo model studies based on PYTHIA 8.107~\cite{PYTHIA} are performed. 
PYTHIA events with high $E_{T}$ jets are generated in three different configurations:\\ 
(i) PyTrue:  PYTHIA isolated jets including all particles except neutrinos.  Jets are  reconstructed using the PYTHIA internal jet finder, PyCell, for the cone algorithm, and $FastJet$ for the sequential recombination algorithm. \\
(ii) PyDet:  PYTHIA isolated jets (parameterized detector response level) reconstructed using the jet algorithms that are also applied to the real data. \\
(iii) PyEmbed:  PyDet that are embedded in a background of real Au+Au 0-10\% central events, with jets reconstructed with  the jet algorithms that are also applied to the real data. \\

In all the real events and simulations, only the highest energy jet per event is selected as the reconstructed jet.  

\subsection{Energy Resolution}
\begin{figure}[h!]
\begin{center}
\resizebox{0.80\columnwidth}{!}{% 
  \includegraphics{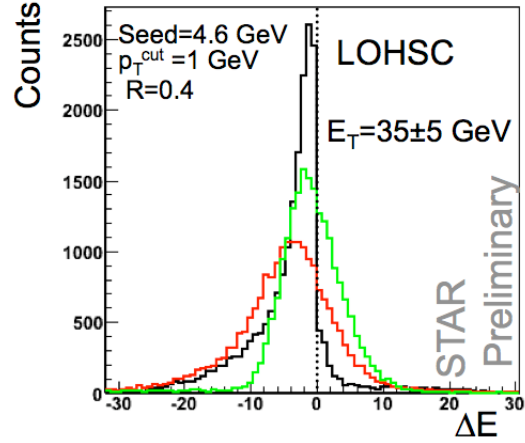} }
  \end{center}
\caption{Distributions showing energy resolution; black   $\rm \Delta E =  E_{PyDet} - E_{PyTrue}$, red $\rm \Delta E = E_{PyEmbed} - E_{PyTrue}$ 
and green  $\rm \Delta  E = E_{PyEmbed} - E_{PyDet}$.}
\label{fig:res}       % Give a unique label
\end{figure}

The energy resolution for jet reconstruction with various algorithms has been studied with isolated jets simulated with PYTHIA \cite{PYTHIA,yichun,mark}.  Figure \ref{fig:res} shows the event by event comparison of PyTrue, PyDet and PyEmbed from LOHSC algorithm. See \cite{me} for comparison of the energy resolution with the $\rm k_{T}$ and Cambridge/Aachen algorithms. Applied cuts and jet energy are specified in the figures. A shift of median due to un-measured particles, primarily neutrons and $K^{0}_{L}$, and the applied  $p_{T}$ cut (hence loss of jet energy), is observed for the  $\rm \Delta E =  E_{PyDet} - E_{PyTrue} $ histogram.  Background  effects are simulated using  PYTHIA jets that are embedded in real $Au+Au$ events. The distribution in Figure~\ref{fig:res} is convoluted with the true jet spectrum to produce the observed jet spectrum.  %A smearing due to background subtraction in Au+Au is seen for all three algorithms for $\rm \Delta  E = E_{PyEmbed} - E_{PyDet}$ histograms. A gain in energy due to undersubtraction of background subtraction is seen as the positive tail of $\rm \Delta E =  E_{PyEmbed} - E_{PyTrue} $ histograms. This effect  is larger in the LOHSC. 
 The effect of the heavy ion background on the jet energy can be seen in the $\rm \Delta E=E_{PyEmbed}-E_{PyTrue}$ distribution.  A positive  $\rm \Delta E$ in this distribution can distort the measured inclusive jet spectrum substantially, increasing the apparent yield at high $E_{T}$ and resulting in a harder spectrum.  A correction to the spectrum must be applied to account for this effect.

\begin{figure}[h!]
\begin{center}
\resizebox{0.80\columnwidth}{!}{% 
  \includegraphics{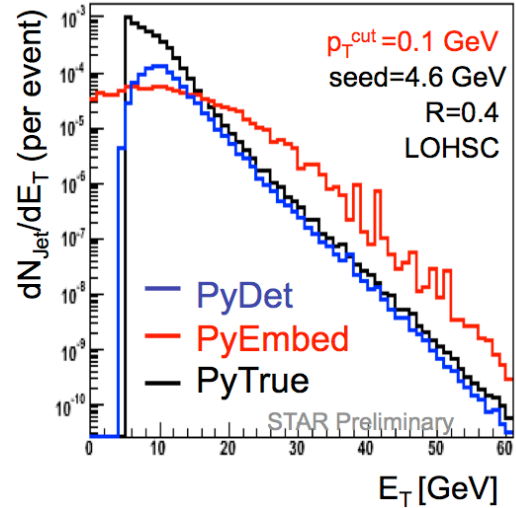} }
  \end{center}
\caption{Inclusive jet spectrum for PyDet, PyEmbed and PyTrue using the LOHSC algorithm. The $p_{T}^{cut}=0.1$ GeV on track momentum and calorimeter cell energy is applied for PyDet and PyEmbed.   Note the lower threshold on generated jet energy $E_{T}^{PyTrue} > 5$ GeV, which affects the reconstructed spectrum up to $E_{T}=20$ GeV.}
\label{fig:reseff}       % Give a unique label
\end{figure}

The influence of energy resolution on jet spectrum can be observed with PyDet, PyEmbed and PyTrue distributions shown in Figure \ref{fig:reseff} for the LOHSC algorithm for $p_{T}^{cut}=0.1$  GeV.  A large difference between PyEmbed and PyDet is observed as expected from the tail at positive $\Delta E$ due to large background in the red distribution in Figure~\ref{fig:res}. When the $\rm p_{T}$ threshold is increased, the background fluctuations are reduced and the enhancement in the spectrum relative to the case without background is reduced to a negligible level \cite{me}. Jet reconstruction in 0-10\% most central Au+Au collisions is similar to that of p+p collisions with a larger $p_{T}^{cut}$ GeV threshold requirement. However, a reduction in the measured jet energy ($\rm p_{T}^{cut}$ dependent bias) is introduced. Similar effects, though smaller in magnitude, are also observed for the $\rm k_{T}$ and Cambridge/Aachen algorithms.

\subsection{Jet Spectra Corrections and Comparisons}

The correction factors for the jet spectrum are estimated using PYTHIA simulated jets embedded in real Au+Au collisions . The $\rm E_{T}$ dependent ratio of PyEmbed to PyTrue is calculated from Figure~\ref{fig:reseff}.  A polynomial function fit to the ratio distribution  is used as a multiplicative correction to the inclusive spectrum. Table~\ref{tab:1} shows the inclusive jet spectrum correction factors for various $p_{T}^{cut}$ values.  For the sequential clustering algorithms, the correction factors are closer to unity.

\begin{table}
\caption{ 
Correction factors for the inclusive jet spectrum, for different reconstruction algorithms and values of $p_{T}^{cut}$. 
The range of values given indicates the correction factor variation from the lowest to highest jet $E_{T}$ shown in the figures.
}
\label{tab:1}       % Give a unique label
\begin{center}
\begin{tabular}{l|l|l|l}
\hline\noalign{\smallskip}
$p_{T}^{cut}$ & LOHSC &  $\rm k_{T}$  & CAMB \\
\noalign{\smallskip}\hline\noalign{\smallskip}
0.1 GeV & 0.2-10 & 1-4 & 2-6 \\
1 GeV & 0.2-1 & 0.7-1 & 1-2\\
2 GeV & 0.2-0.3 & 0.5-1 & 0.5-1\\
\noalign{\smallskip}\hline
\end{tabular}
\end{center}
\end{table}

\begin{figure}[h!]
\begin{center}
\resizebox{0.4\textwidth}{!}{%
	\includegraphics{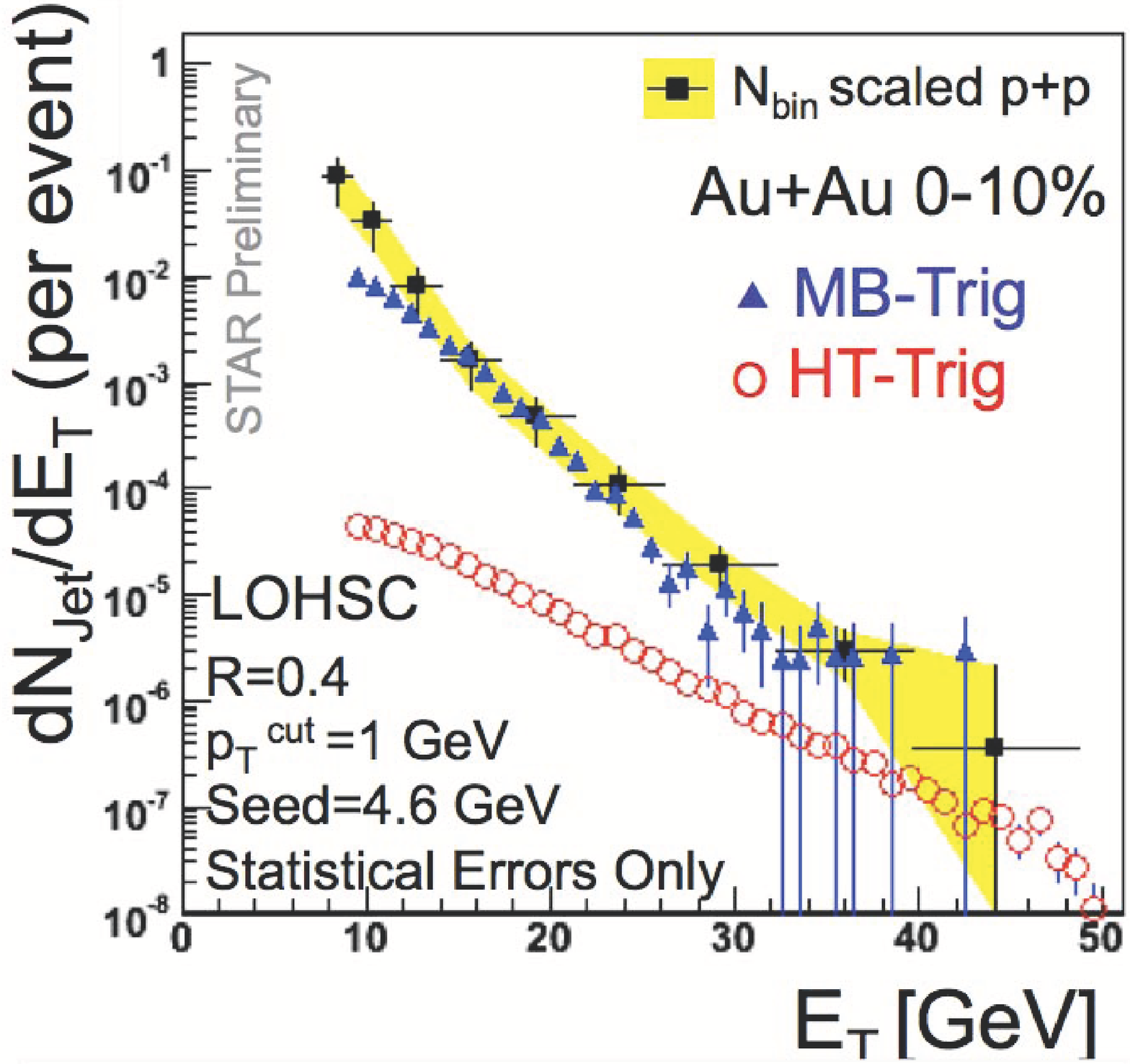} 
}	
\end{center}
\caption{Jet yield per event vs $E_{T}$ for 0-10\% central $Au+Au$ collisions, compared to the distribution from $p+p$ collisions scaled by  $\rm N_{Binary}$ \cite{starpp}.   
Triangle symbols are from MB-Trig and corrected for efficiency, acceptance and energy resolution. Open circles are from HT-Trig and not corrected for trigger bias. Only statistical error bars are shown for the $Au+Au$ data. Solid black squares are the distribution from $p+p$ collisions, scaled by $N_{Binary}$.  The yellow band represents the systematic uncertainty of the $p+p$ measurement. }
\label{fig:spectra}       % Give a unique label
\end{figure}

The corrected inclusive jet spectrum for the LOHSC algorithm for the $p_{T}^{cut}=1$ GeV is presented in Figure~\ref{fig:spectra}.  The $p_{T}^{cut}=1$ GeV  is selected as it corresponds to correction factors close to unity. At this $p_{T}^{cut}$ the competing effects of energy loss due to momentum threshold cut and the kick up in the jet spectrum due to the positive tail of energy resolution cancel each other. The solid triangles are for the MB-Trig data set and are corrected for resolution, acceptance and efficiency.

In order to assess the biases in the jet spectrum reconstructed in central Au+Au collisions, we compare to the spectrum measured in p+p collisions. To account for nuclear geometric effects we scale the p+p spectrum by $\rm N_{Binary}$, the number of binary nucleon+nucleon collisions equivalent to a central Au+Au collisions, as calculated by a Glauber model \cite{glauber}. The cross section for hard processes is expected to scale with $\rm N_{Binary}$ if no nuclear effects are present. In the case of jet reconstruction, $\rm N_{Binary}$ scaling is expected if the reconstruction is unbiased, i.e. the jet energy is recovered independent of the particular mode of fragmentation, even in the presence of strong jet quenching. The $N_{Binary}$ scaled  jet spectrum from $p+p$ collisions is shown in solid squares \cite{starpp}.   The yellow band represents the systematic uncertainty of the $p+p$ jet measurement.  Heavy ion jet spectrum is observed to agree with $N_{Binary}$ scaled p+p measurement within the $\sim 50 \%$ systematic uncertainty of the normalization.

The open circles  in Figure~\ref{fig:spectra} show the uncorrected jet spectrum from HT-Trig data which is substantially lower than the corrected MB-Trig spectrum.  The correction factors of energy resolution, efficiency and acceptance for the HT-Trig are  expected to be small.  A large trigger bias due to the additional 7.5 GeV energy deposition in EMCAL requirement for HT-Trig relative to MB-Trig is seen to persist at least to 30 GeV.  Further statistics of MB-Trig data is needed to assess the bias at high $p_{T}$. %and the correction needed for the jet spectrum from HT-Trig. When the rest of the MB-Trig data is processed from year 7, these corrections for the spectrum will be estimated.  

\begin{figure}[h!]
\begin{center}
\resizebox{0.40\textwidth}{!}{%
	\includegraphics{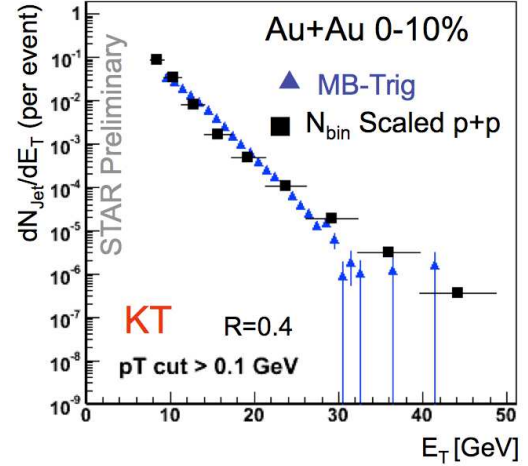} 
}	
\end{center}
\caption{Jet yield per event vs $E_{T}$ for 0-10\% central $Au+Au$ collisions obtained by the $\rm k_{T}$ algorithm. The distribution from $p+p$ collisions are scaled by  $\rm N_{Binary}$ \cite{starpp}.  Triangle symbols are from MB-Trig and corrected for efficiency, acceptance and energy resolution. Only statistical error bars are shown for the $Au+Au$ data. Solid black squares are the distribution from $p+p$ collisions, scaled by $N_{Binary}$. The systematic uncertainty of the $p+p$ jet spectra normalization is  $\sim 50 \%$. }
\label{fig:kt}       % Give a unique label
\end{figure}

\begin{figure}[h!]
\begin{center}
\resizebox{0.40\textwidth}{!}{%
	\includegraphics{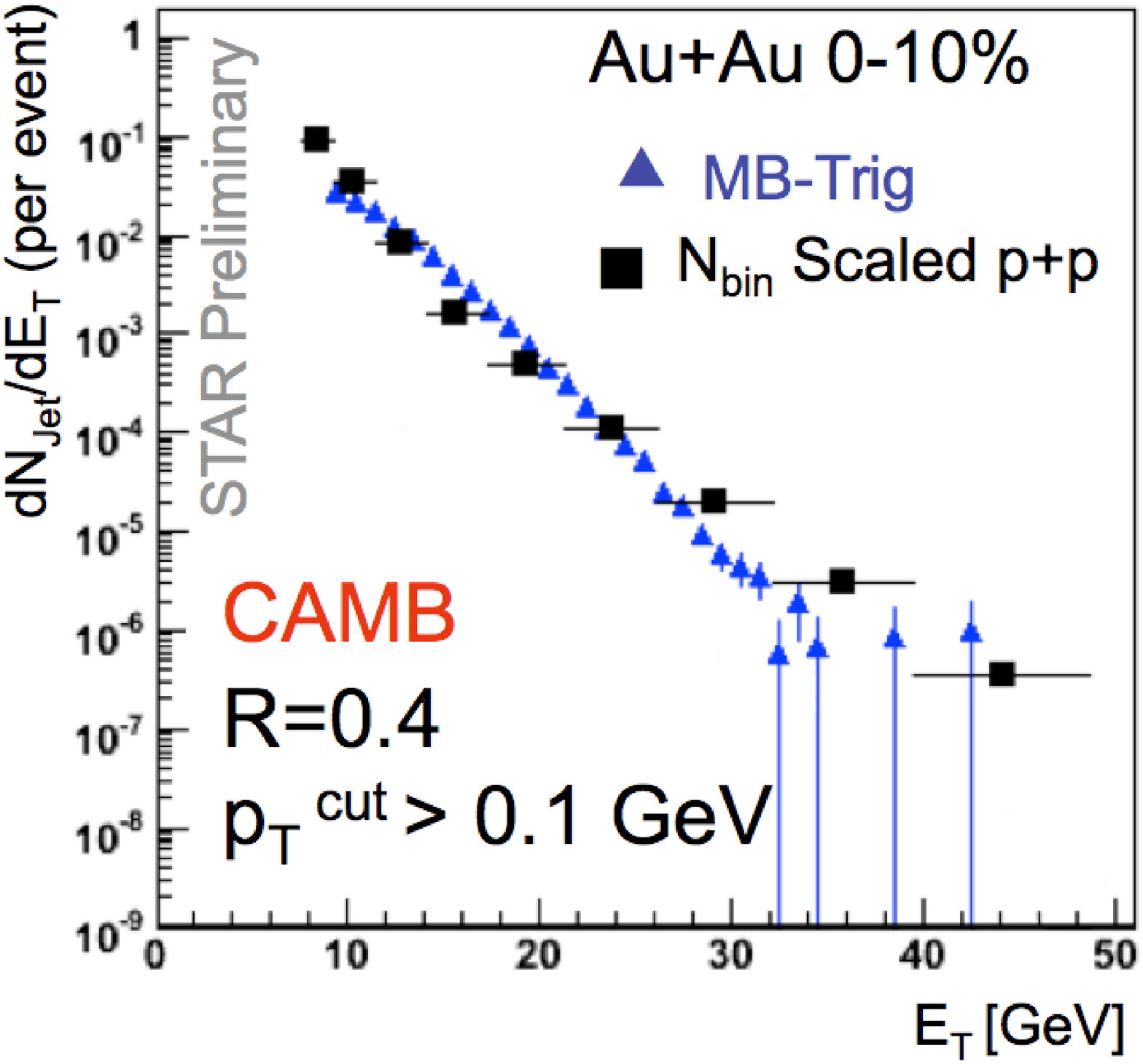} 
}	
\end{center}
\caption{Jet yield per event vs $E_{T}$ for 0-10\% central $Au+Au$ collisions obtained by the Cambridge/Aachen  algorithm. The distribution from $p+p$ collisions are scaled by  $\rm N_{Binary}$ \cite{starpp}.  Triangle symbols are from MB-Trig and corrected for efficiency, acceptance and energy resolution.  Only statistical error bars are shown for the $Au+Au$ data. Solid black squares are the distribution from $p+p$ collisions, scaled by $N_{Binary}$. The systematic uncertainty of the $p+p$ jet spectra normalization is  $\sim 50 \%$.}
\label{fig:camb}       % Give a unique label
\end{figure}

\begin{figure}[h!]
\begin{center}
\resizebox{0.40\textwidth}{!}{%
	\includegraphics{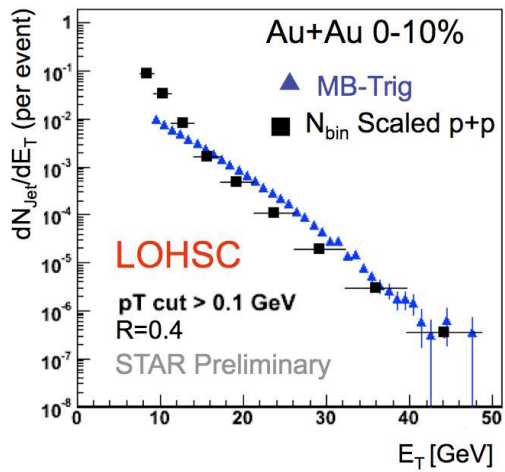} 
}	
\end{center}
\caption{Jet yield per event vs $E_{T}$ for 0-10\% central $Au+Au$ collisions obtained by the LOHSC algorithm. The distribution from $p+p$ collisions are scaled by  $\rm N_{Binary}$ \cite{starpp}.   
Triangle symbols are from MB-Trig and corrected for efficiency, acceptance and energy resolution.  Only statistical error bars are shown for the $Au+Au$ data. Solid black squares are the distribution from $p+p$ collisions, scaled by $N_{Binary}$. The systematic uncertainty of the $p+p$ jet spectra normalization is  $\sim 50 \%$.}
\label{fig:cone}       % Give a unique label
\end{figure}

Figures~\ref{fig:kt},\ref{fig:camb},\ref{fig:cone} show the comparison of inclusive jet spectra from the MB-Trig $Au+Au$ data and the $N_{Binary}$ scaled p+p for the  $p_{T}=0.1$ GeV threshold cut for $\rm k_{T}$, Cambrdige/Aachen and LOHSC algorithms. While the agreement between $N_{Binary}$ scaled p+p and MB-Trig measurement is good for $p_{T}^{cut}=0.1$ GeV, it is also seen to be  poorer with  the larger $p_{T}$ threshold cut \cite{me}. This suggests that $p_{T}^{cut}$ introduces biases which are not fully corrected using PYTHIA as the fragmentation model, and/or may be an indication of modified fragmentation due to quenching.

\section{Summary}

The full reconstruction of jets in 0-10\% most central heavy ion collisions at RHIC energies is presented.   Systematics of the underlying heavy ion background subtraction is studied utilizing various algorithms and consideration of the jet area. The $\rm N_{Binary}$ scaling is observed for the least-biased cuts with the given $\sim 50\%$ systematic uncertainty of the $p+p$ jet spectrum measurement.   
An unbiased jet reconstruction in heavy ion collisions with MB-Trig data appears to be feasible. However, spectrum corrections are currently based on model calculations using PYTHIA fragmentation. This aspect, together with background subtraction techniques, spectrum variations due to cuts and reconstruction algorithms, must be investigated further in order to assess the systematic uncertainties of this measurement.

The first heavy ion run at LHC is expected in late 2009. The heavy ion background is predicted to be larger at the LHC than at RHIC, but there will be copious production of very energetic jets, well above background \cite{peter}. The large kinematic reach at the LHC may provide sufficient lever-arm to map out the QCD evolution  of jet quenching \cite{solan}.  The comparison of  full jet measurements in the different physical systems generated at RHIC and the LHC will provide unique and crucial insights into our understanding of jet quenching and the nature of hot QCD matter.

% BibTeX users please use
% \bibliographystyle{}
% \bibliography{}

\end{document}